\documentclass[a4paper, 11pt]{article}
\usepackage[top=18truemm,bottom=25truemm,left=25truemm,right=25truemm]{geometry}

\usepackage{indentfirst}

\makeatletter
\def\section{\@startsection {section}{1}{\z@}{-3.5ex plus -1ex minus -.2ex}{2.3 ex plus .2ex}  {\large\bf}}
\makeatother
\usepackage{secdot}

\makeatletter
\def\subsection{\@startsection {subsection}{1}{\z@}{-3.5ex plus -1ex minus -.2ex}{2.3 ex plus .2ex}{\small\bf}}
\makeatother

\usepackage{amsmath, amssymb}
\usepackage{siunitx}
\usepackage{graphicx}
\usepackage[labelsep=period, figurename=Fig. ,tablename=Table]{caption}
\usepackage{subcaption}

\usepackage[numbers,sort&compress]{natbib}
\usepackage{url}

\usepackage[T1]{fontenc}
\usepackage{color}

\title{Quantifying defensive pressure on the ball carrier in soccer based on minimum arrival time}

\usepackage{authblk}
\usepackage[hidelinks]{hyperref}

\providecommand{\DOIprefix}{}
\renewcommand{\DOIprefix}{}
\providecommand{\doi}[1]{}
\renewcommand{\doi}[1]{\href{https://doi.org/#1}{\nolinkurl{doi:#1}}}

\author[1]{Takuma Narizuka\thanks{{\it E-mail address}: narizuka.takuma@gmail.com\\
}}
\author[2]{Ikuya Sakamoto}
\author[3]{Ken Yamamoto}
\author[4]{Yoshihiro Yamazaki}
\affil[1]{Faculty of Data Science, Rissho University, Kumagaya, Saitama 360-0194, Japan}
\affil[2]{Department of Physics, Faculty of Science and Engineering, Chuo University, Bunkyo, Tokyo 112-8551, Japan}
\affil[3]{Faculty of Science, University of the Ryukyus, Nishihara, Okinawa 903-0213, Japan}
\affil[4]{Department of Physics, School of Advanced Science and Engineering, Waseda University, Shinjuku, Tokyo 169-8555, Japan}
\date{}

\providecommand{\keywords}[1]
{
  \noindent{\footnotesize\textbf{\textit{Keywords---}} #1}
  \small
}

\begin{document}
	\maketitle
	\baselineskip 14pt

\begin{abstract}
Defensive pressure on the ball carrier is a fundamental component of soccer tactics.
Existing pressure measures often involve additional modeling assumptions, which may reduce interpretability.
In this study, we quantify defensive pressure as the opponent minimum arrival time to the ball-carrier location, computed from a physics-based motion model.
Using synchronized event and tracking data from all 306 matches in the top division of the Japan Professional Football League during the 2023 season,
we analyze the statistical characteristics and temporal evolution of this quantity during ball-possession intervals.
The results show that the opponent minimum arrival time tends to decrease during possession and to increase again at the start of the next possession after ball release.
We also find that possessions starting under stronger defensive pressure tend to yield smaller ball progression, and that, for intentional open-play passes, possessions ending under stronger pressure are more likely to be lost.
These findings indicate that minimum arrival time provides an interpretable and physically grounded measure of immediate defensive pressure on the ball carrier.
The proposed framework provides a simple and interpretable baseline for quantifying pressing dynamics from tracking data.
\end{abstract}


\section{Introduction}
Soccer (football) is a complex team sport in which 22 players interact continuously in a large field.
Recent advances in synchronized event and tracking data~\cite{Bassek2025} have enabled quantitative analyses of player movements and tactical behaviors at high temporal resolution~\cite{Gudmundsson2017, Wang2024}.
From a statistical-physics perspective, soccer can be viewed as a system of interacting self-propelled agents, where local interactions between attackers and defenders are reflected in their relative positions, velocities, and arrival times.
One key aspect of soccer tactics is spatial control~\cite{Fernandez2018,Narizuka2021}.
Attacking players attempt to create space and provide passing options, whereas defenders aim to restrict these opportunities by applying pressure on the ball carrier.
Such local interactions strongly influence subsequent on-ball outcomes, including ball progression, possession retention, pass success, and shot creation.
Motivated by the importance of defensive pressure, a number of approaches have been proposed to quantify it.
Early work described pressure using geometric relationships such as relative distances and angles between players~\cite{Andrienko2017}, and such formulations were later adapted to analyze pressure in different game contexts~\cite{Herold2022,Forcher2024}.
Other studies have focused on detecting pressing situations using supervised learning or rule-based approaches~\cite{Bauer2021,Peters2026}, or on evaluating the effectiveness of pressing strategies from a risk--reward perspective~\cite{Merckx2021}.
More recently, time-to-intercept measures derived from motion models have been used to construct probabilistic representations of defensive pressure~\cite{Spearman2017,Bekkers2025}.
These previous studies have substantially advanced the quantitative analysis of pressing.
At the same time, existing approaches are designed either to detect specific pressing events or to transform arrival-time information into more elaborate pressure scores.
Such formulations are useful for particular applications, but they may also introduce additional assumptions that make interpretation less direct.
From this perspective, minimum arrival time itself provides a simple and interpretable measure of immediate defensive pressure.
It directly represents how quickly a defender can reach the ball-carrier location under a specified motion model~\cite{Fujimura2005,Narizuka2023}.
Because it is defined directly from player kinematics, it enables pressure to be quantified with fewer additional assumptions than more elaborate pressure scores.
This makes it suitable as a baseline quantity for describing pressing dynamics in an interpretable manner.
In this study, we analyze defensive pressure on the ball carrier using the opponent minimum arrival time derived from a physics-based motion model.
Using synchronized tracking and event data from 306 J1 League matches, we investigate the statistical characteristics and temporal evolution of this quantity during ball-possession intervals.
We further examine how the opponent minimum arrival time is associated with two key possession outcomes, namely ball progression and possession loss.
Overall, this study shows that the opponent minimum arrival time can be interpreted as a local first-arrival-time measure of defensive interaction around the ball carrier and provides an interpretable baseline for quantifying immediate defensive pressure from tracking data.

\section{Method}

\subsection{Data and preprocessing}
We used synchronized event and tracking data from all 306 matches in the top division of the Japan Professional Football League (J1 League) during the 2023 season.
The tracking data consist of the absolute coordinates $ (x, y) $ for all players, recorded at 25 frames per second, while the event data contain time-stamped on-ball actions such as passes, shots, and fouls.
The datasets were provided by DataStadium Inc., Japan~\cite{DataStadium}, and we obtained explicit permission to use them in this study.
Following a preprocessing procedure used in a previous study~\cite{Brink2023},
measurement noise in the tracking data was reduced using the Savitzky--Golay
filter and cubic spline interpolation.
Player velocities were then computed from the smoothed trajectories.
Player locations were represented in a pitch-fixed coordinate system, and attacking direction was standardized across halves.

\subsection{Motion model and minimum arrival time}\label{subsec:method_motion_model}
To quantify spatial pressure in terms of the time required to reach a location, we computed the minimum arrival time using a simplified physics-based motion model.
Specifically, we used the Fujimura--Sugihara model~\cite{Fujimura2005}, in which the motion of player $ p $ is governed by
\begin{equation}
  m \frac{d^2 \vec{x}_{p}(t)}{dt^2} = F \vec{n} - k \frac{d \vec{x}_{p}(t)}{dt},
  \label{eq:fujimura-sugihara}
\end{equation}
where $ \vec{x}_{p}(t) $ is the position vector of player $ p $ at time $ t $, $ m $ is the player mass, $ F\vec{n} $ is a constant driving force in direction $ \vec{n} $, and $ k $ is a viscous resistance coefficient.
Under the assumption that the direction of the driving force $ \vec{n} $ can change freely, the set of reachable locations after time $ t $ forms a circle with center $ \vec{c}_p(t) $ and radius $ r_p(t) $:
\begin{align}
  \vec{x}_{p}(t)
  &= \vec{x}_{p}(0) + \frac{1 - \exp(-\alpha t)}{\alpha} \vec{v}_{p}(0)
  + V_{\max} \left(t - \frac{1 - \exp(-\alpha t)}{\alpha}\right)\vec{n} \nonumber \\
  &= \vec{c}_{p}(t) + r_{p}(t) \vec{n},
  \label{eq:fujimura-sugihara_solution}
\end{align}
where
\begin{equation}
\begin{aligned}
  \vec{c}_{p}(t)
  &= \vec{x}_{p}(0) + \frac{1 - \exp(-\alpha t)}{\alpha} \vec{v}_{p}(0), \\
  r_{p}(t)
  &= V_{\max} \left(t - \frac{1 - \exp(-\alpha t)}{\alpha}\right).
\end{aligned}
\label{eq:center_radius}
\end{equation}
Here, $ \alpha = k/m $ and $ V_{\max} = F/k $ are kinetic parameters.
Following our previous validation study of the Fujimura--Sugihara model under soccer players' sprint conditions~\cite{Narizuka2023}, we used \(\alpha = \qty{1.0}{\per\second}\) and \(V_{\max} = \qty{10.0}{\metre\per\second}\).
Given a target location $ \vec{x} $, the arrival time $ \tau_{p}(\vec{x}, t) $ for player $ p $ is computed by numerically solving
\begin{equation}
  r_{p}(\tau_{p}) = \lVert \vec{x} - \vec{c}_{p}(\tau_{p}) \rVert.
  \label{eq:arrival_time_equation}
\end{equation}
The minimum arrival time to location $ \vec{x} $ at time $ t $ is then defined as the minimum over all candidate players.

\subsection{Pressure on the ball carrier}
\label{subsec:method_pressure}
\begin{figure}[htbp]
  \centering
  \includegraphics[width=0.4\linewidth]{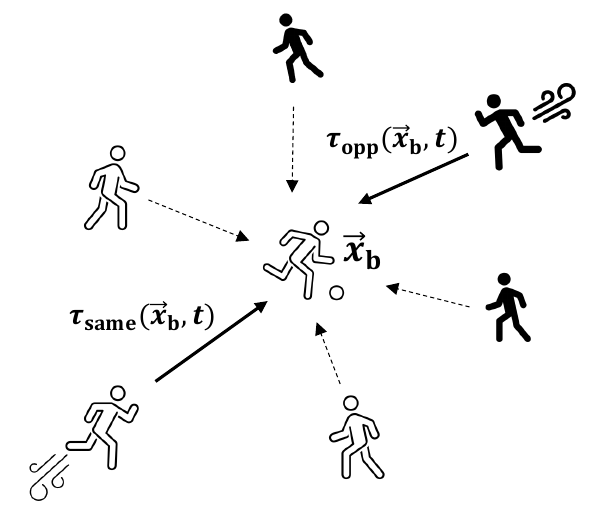}
  \caption{
  Schematic illustration of the opponent and teammate minimum arrival times to the ball-carrier location, $\tau_{\mathrm{opp}}(\vec{x}_{\mathrm{b}},t)$ and $\tau_{\mathrm{same}}(\vec{x}_{\mathrm{b}},t)$, respectively.
  }
  \label{fig:tau_player_def}
\end{figure}

To evaluate defensive pressure on the ball carrier, we identified ball-possession intervals from the synchronized event and tracking data.
A ball-possession interval was defined as a continuous interval in the synchronized event-tracking record during which a single player was identified as controlling the ball.
We denote the possession start time by $t_{\mathrm{get}}$ and the ball-release time by $t_{\mathrm{rel}}$.
Intervals with $t_{\mathrm{rel}} > t_{\mathrm{get}}$ were classified as possession plays, whereas those with $t_{\mathrm{rel}} = t_{\mathrm{get}}$ were classified as direct plays.
Goalkeeper possessions were excluded from all analyses because their tactical role and surrounding pressure structure differ substantially from those of outfield players.
Let $\vec{x}_{\mathrm{b}}$ denote the ball-carrier location at time $ t $.
Using the motion model in Subsection~\ref{subsec:method_motion_model}, we computed the opponent and teammate minimum arrival times to the ball-carrier location as
\[
  \tau_{\mathrm{opp}}(\vec{x}_{\mathrm{b}},t)
  =
  \min_{p\in\mathrm{opponents}}\tau_p(\vec{x}_{\mathrm{b}}, t),
  \qquad
  \tau_{\mathrm{same}}(\vec{x}_{\mathrm{b}},t)
  =
  \min_{p\in\mathrm{teammates}}\tau_p(\vec{x}_{\mathrm{b}}, t).
\]
The ball carrier was excluded from the calculation of \(\tau_{\mathrm{same}}\).
In particular, we extracted $\tau_{\mathrm{opp}}(\vec{x}_{\mathrm{b}},t_{\mathrm{get}})$ and $\tau_{\mathrm{opp}}(\vec{x}_{\mathrm{b}},t_{\mathrm{rel}})$ at possession start and ball release, respectively, and used them as measures of immediate defensive pressure on the ball carrier.
A schematic illustration of $\tau_{\mathrm{opp}}(\vec{x}_{\mathrm{b}},t)$ and $\tau_{\mathrm{same}}(\vec{x}_{\mathrm{b}},t)$ is shown in Fig.~\ref{fig:tau_player_def}.
For the possession-loss analysis in Subsection~\ref{subsec:result_loss_analysis}, we further restricted the release actions to intentional teammate-directed open-play passes, namely ordinary passes and through passes as labeled in the event data.
Ball loss was defined as the situation in which possession switched to the opponent or play stopped after the action.
The numbers of analyzed possessions and passes are reported for the ball-progression and possession-loss analyses.

\section{Results}

\subsection{Distribution of minimum arrival times to the ball carrier}
\begin{figure}[htbp]
  \centering
  \includegraphics[width=0.5\linewidth]{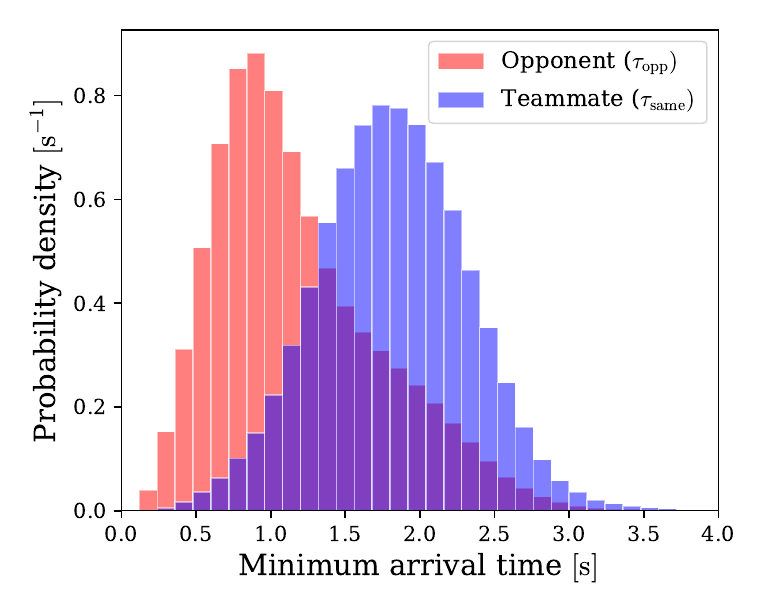}
  \caption{
    Distributions of the opponent and teammate minimum arrival times, $\tau_{\mathrm{opp}}(\vec{x}_{\mathrm{b}},t)$ and $\tau_{\mathrm{same}}(\vec{x}_{\mathrm{b}},t)$, to the ball-carrier location during ball-possession intervals across 306 matches.
  }
  \label{fig:hist_tau_same-opp}
\end{figure}

We first examined the overall distributions of the minimum arrival times to the ball carrier.
Figure~\ref{fig:hist_tau_same-opp} compares the opponent and teammate minimum arrival times, $ \tau_{\mathrm{opp}}(\vec{x}_{\mathrm{b}},t) $ and $\tau_{\mathrm{same}}(\vec{x}_{\mathrm{b}},t) $, during ball-possession intervals across 306 matches.
The distribution of $\tau_{\mathrm{same}}(\vec{x}_{\mathrm{b}},t) $ is approximately symmetric, whereas $\tau_{\mathrm{opp}}(\vec{x}_{\mathrm{b}},t) $ is clearly concentrated at shorter times.
This asymmetry indicates that opponents tend to reach the ball carrier more quickly than teammates.
This is consistent with their different tactical roles: opponents apply pressure to the ball carrier, whereas teammates are often positioned to maintain their formation while providing support.
Since the main focus of this study is defensive pressure on the ball carrier, the following analyses concentrate on the temporal behavior and consequences of the opponent minimum arrival time to the ball-carrier location.

\subsection{Temporal evolution of defensive pressure during possession}
\begin{figure}[htbp]
  \centering
  \includegraphics[width=\linewidth]{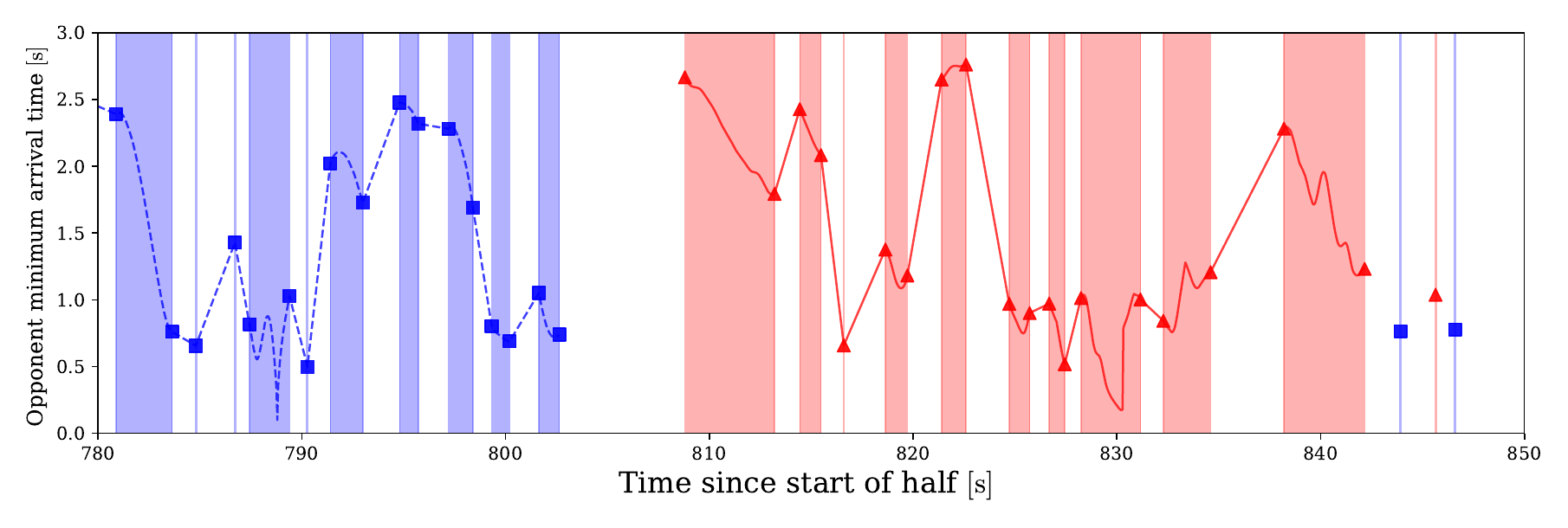}
  \caption{
    Typical time series of the opponent minimum arrival time to the ball-carrier location,
    \(\tau_{\mathrm{opp}}(\vec{x}_{\mathrm{b}}, t)\), from Matchweek 2 of the 2023 J1 League season between F.C. Tokyo and Urawa Red Diamonds.
    F.C. Tokyo possessions are shown by solid lines with triangle markers, whereas Urawa Red Diamonds possessions are shown by dashed lines with square markers.
    The shaded bands indicate individual ball-possession intervals.
  }
  \label{fig:time_series}
\end{figure}

\begin{figure}[htbp]
  \centering
  \begin{subfigure}[t]{0.48\linewidth}
    \centering
    \includegraphics[width=\linewidth]{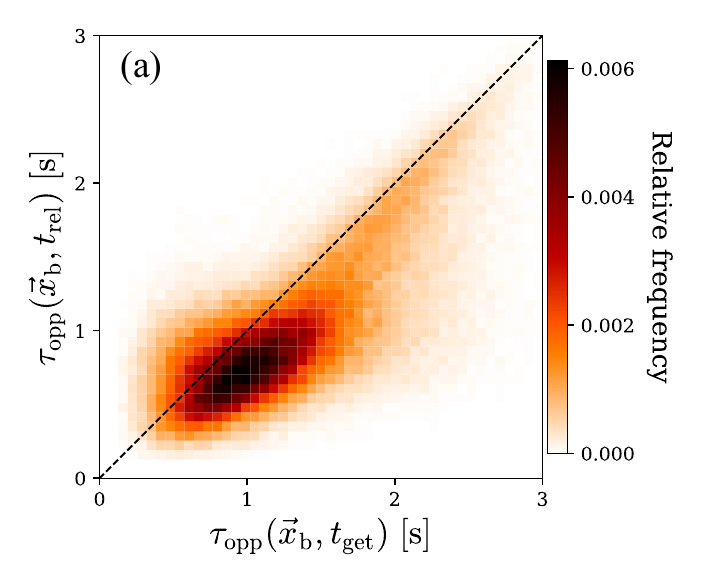}
  \end{subfigure}
  \hfill
  \begin{subfigure}[t]{0.48\linewidth}
    \centering
    \includegraphics[width=\linewidth]{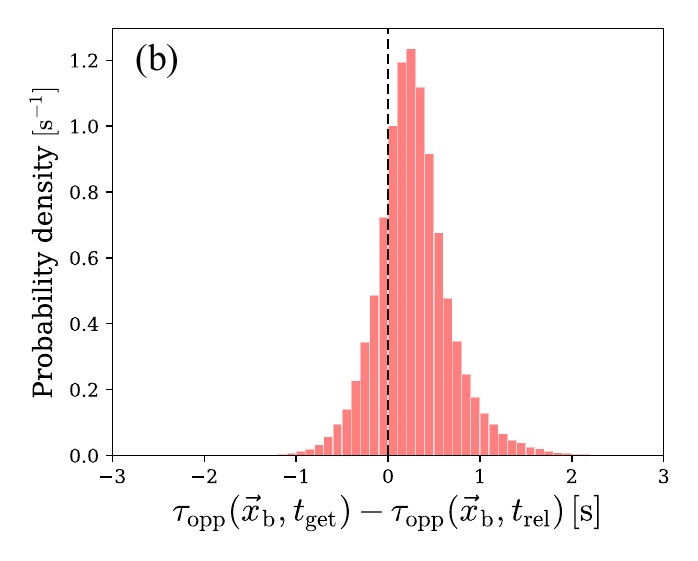}
  \end{subfigure}
  \caption{
      Temporal change in the opponent minimum arrival time during possession.
      (a) Normalized two-dimensional histogram of $\tau_{\mathrm{opp}}(\vec{x}_{\mathrm{b}},t_{\mathrm{get}})$ and $\tau_{\mathrm{opp}}(\vec{x}_{\mathrm{b}},t_{\mathrm{rel}})$ for all possessions excluding direct plays; the dashed line indicates equality.
      (b) Distribution of the difference $\tau_{\mathrm{opp}}(\vec{x}_{\mathrm{b}},t_{\mathrm{get}})-\tau_{\mathrm{opp}}(\vec{x}_{\mathrm{b}},t_{\mathrm{rel}})$.
  }
  \label{fig:tau_get-tau_rel}
\end{figure}

We next investigated how defensive pressure evolves during ball possession.
Figure~\ref{fig:time_series} shows a typical time series.
The two teams are distinguished by line style and marker shape, and the shaded bands indicate individual ball-possession intervals.
As shown in Fig.~\ref{fig:time_series}, the opponent minimum arrival time, $\tau_{\mathrm{opp}}(\vec{x}_{\mathrm{b}}, t)$, tends to decrease while a player is in possession and to increase again at the start of the next possession after ball release.
These observations suggest that defensive pressure gradually builds up on the current ball carrier and is partially relieved when the next possession begins.
To quantify these tendencies, we compared $ \tau_{\mathrm{opp}}(\vec{x}_{\mathrm{b}},t_{\mathrm{get}}) $ and $ \tau_{\mathrm{opp}}(\vec{x}_{\mathrm{b}},t_{\mathrm{rel}}) $ for all possessions excluding direct plays.
Figure~\ref{fig:tau_get-tau_rel}(a) shows the normalized two-dimensional histogram of $ \tau_{\mathrm{opp}}(\vec{x}_{\mathrm{b}},t_{\mathrm{get}}) $ and $ \tau_{\mathrm{opp}}(\vec{x}_{\mathrm{b}},t_{\mathrm{rel}}) $.
Most observations lie below the identity line, indicating that $ \tau_{\mathrm{opp}} $ is typically smaller at ball release than at possession start.
Figure~\ref{fig:tau_get-tau_rel}(b) shows the distribution of the difference $ \tau_{\mathrm{opp}}(\vec{x}_{\mathrm{b}},t_{\mathrm{get}})-\tau_{\mathrm{opp}}(\vec{x}_{\mathrm{b}},t_{\mathrm{rel}}) $, which is shifted toward positive values.
The mean difference was \qty{0.276}{\second} and the standard deviation was \qty{0.400}{\second}, confirming that defensive pressure tends to increase during possession.

\begin{figure}[htbp]
  \centering
  \begin{subfigure}[t]{0.48\linewidth}
    \centering
    \includegraphics[width=\linewidth]{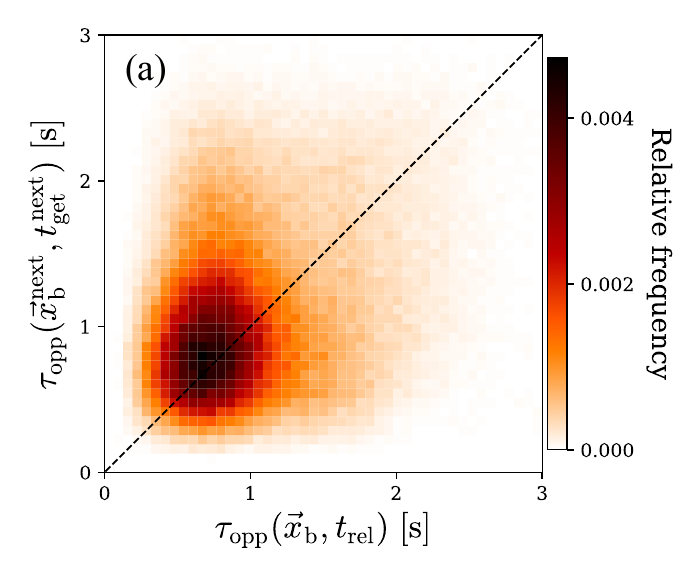}
  \end{subfigure}
  \hfill
  \begin{subfigure}[t]{0.48\linewidth}
    \centering
    \includegraphics[width=\linewidth]{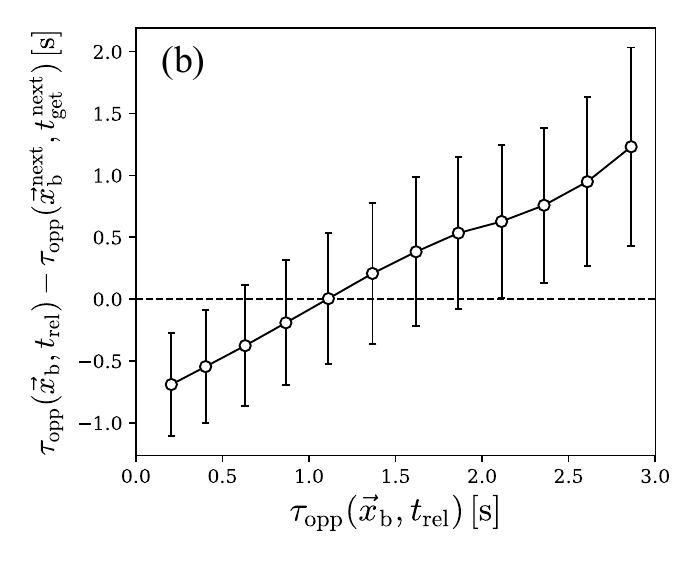}
  \end{subfigure}
  \caption{
    Change in the opponent minimum arrival time after ball release.
    (a) Normalized two-dimensional histogram of $\tau_{\mathrm{opp}}(\vec{x}_{\mathrm{b}}, t_{\mathrm{rel}})$ in one possession and $\tau_{\mathrm{opp}}(\vec{x}_{\mathrm{b}}^{\,\mathrm{next}}, t_{\mathrm{get}}^{\,\mathrm{next}})$ at the start of the next possession; the dashed line indicates equality.
    (b) Mean value of $\tau_{\mathrm{opp}}(\vec{x}_{\mathrm{b}}, t_{\mathrm{rel}})-\tau_{\mathrm{opp}}(\vec{x}_{\mathrm{b}}^{\,\mathrm{next}}, t_{\mathrm{get}}^{\,\mathrm{next}})$, plotted as a function of $\tau_{\mathrm{opp}}(\vec{x}_{\mathrm{b}}, t_{\mathrm{rel}})$.
    Error bars indicate the standard deviation within each bin.
    }
  \label{fig:tau_rel-next_tau_get}
\end{figure}

We then examined whether this increase in defensive pressure is followed by a reduction at the start of the next possession after ball release.
Here, $\vec{x}_{\mathrm{b}}^{\,\mathrm{next}}$ and $t_{\mathrm{get}}^{\,\mathrm{next}}$ denote the ball-carrier location and possession-start time in the next possession, respectively.
Figure~\ref{fig:tau_rel-next_tau_get}(a) compares $\tau_{\mathrm{opp}}(\vec{x}_{\mathrm{b}}, t_{\mathrm{rel}})$ in one possession with $\tau_{\mathrm{opp}}(\vec{x}_{\mathrm{b}}^{\,\mathrm{next}}, t_{\mathrm{get}}^{\,\mathrm{next}})$ at the start of the next possession.
Most observations lie above the identity line, indicating that the opponent minimum arrival time tends to be larger at the start of the next possession than at the end of the previous one.
Note that this comparison does not track the same player over time.
Rather, it compares the immediate defensive pressure on the ball carrier at the end of one possession with that on the ball carrier at the start of the subsequent possession.
To examine how defensive pressure changes across consecutive possessions, we analyzed the difference between $\tau_{\mathrm{opp}}(\vec{x}_{\mathrm{b}}, t_{\mathrm{rel}})$ in one possession and $\tau_{\mathrm{opp}}(\vec{x}_{\mathrm{b}}^{\,\mathrm{next}}, t_{\mathrm{get}}^{\,\mathrm{next}})$ at the start of the next possession as a function of $\tau_{\mathrm{opp}}(\vec{x}_{\mathrm{b}}, t_{\mathrm{rel}})$.
Figure~\ref{fig:tau_rel-next_tau_get}(b) shows that this difference varies systematically with $\tau_{\mathrm{opp}}(\vec{x}_{\mathrm{b}}, t_{\mathrm{rel}})$.
In particular, when the current possession ends under stronger pressure, the next possession tends to begin under lower immediate pressure relative to the end of the previous one.
These results suggest that the ball carrier in the next possession tends to face lower immediate defensive pressure than the player who released the ball in the previous possession.

\subsection{Association between defensive pressure and ball progression}
\label{subsec:ball_progression}
\begin{figure}[htbp]
  \centering
  \begin{subfigure}[t]{0.48\linewidth}
    \centering
    \includegraphics[width=\linewidth]{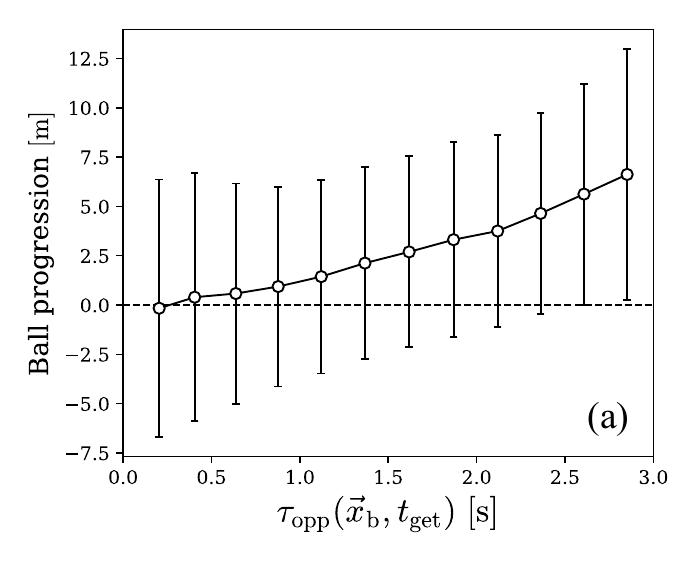}
  \end{subfigure}
  \hfill
  \begin{subfigure}[t]{0.48\linewidth}
    \centering
    \includegraphics[width=\linewidth]{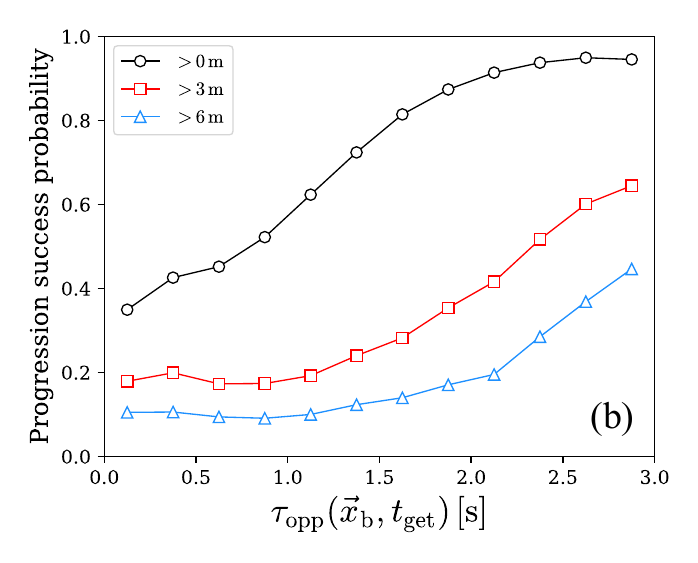}
  \end{subfigure}
  \caption{
    Association between defensive pressure and ball progression.
    (a) Mean ball progression as a function of the opponent minimum arrival time at possession start. Error bars indicate the standard deviation within each bin.
    (b) Progression success probability as a function of the opponent minimum arrival time at possession start. The curves indicate the probability that ball progression exceeds \qty{0}{\metre}, \qty{3}{\metre}, or \qty{6}{\metre}.
  }
  \label{fig:ball_progression}
\end{figure}

We next examined how defensive pressure is associated with ball progression during a possession.
In particular, we analyzed how the opponent minimum arrival time at possession start, $\tau_{\mathrm{opp}}(\vec{x}_{\mathrm{b}}, t_{\mathrm{get}})$, is related to subsequent ball progression.
Here, ball progression was defined as the decrease in the ball carrier's distance to the opponent goal from possession start to end.
This analysis included $N=167735$ ball-possession intervals after excluding goalkeeper possessions and direct plays, corresponding to approximately 548 intervals per match on average.
Figure~\ref{fig:ball_progression}(a) shows that possessions starting under stronger defensive pressure tend to result in smaller ball progression.
Figure~\ref{fig:ball_progression}(b) shows the probability that ball progression exceeds \qty{0}{\metre}, \qty{3}{\metre}, or \qty{6}{\metre} as a function of $\tau_{\mathrm{opp}}(\vec{x}_{\mathrm{b}}, t_{\mathrm{get}})$.
Overall, these results suggest that stronger immediate defensive pressure is associated with reduced ball progression during possession.
One possible confounding factor in this relationship is field location.
Possessions starting closer to the opponent goal tend to have less room for further ball progression and may face stronger defensive pressure because defenders are more densely positioned near the goal.
To examine whether the association in Fig.~\ref{fig:ball_progression}(a) simply reflects such field-location effects, we repeated the analysis after dividing possession plays into three zones according to the ball carrier's distance to the opponent goal at possession start.
The three zones were defined using thresholds of \qty{35}{\metre} and \qty{70}{\metre}.
The same qualitative relationship between \(\tau_{\mathrm{opp}}(\vec{x}_{\mathrm{b}}, t_{\mathrm{get}})\) and ball progression was observed in all three zones (Supplementary Fig.~S1), indicating that the association between stronger immediate defensive pressure and smaller ball progression is not explained solely by field location.

\subsection{Association between defensive pressure and possession loss}
\label{subsec:result_loss_analysis}
In addition to constraining ball progression, defensive pressure at ball release may also increase the risk of losing possession.
To examine this possibility, we analyzed intentional teammate-directed open-play passes, as defined in Subsection~\ref{subsec:method_pressure}.
The analysis included $ N=237596 $ passes in total (approximately 777 per match on average), consisting of $ N=153164 $ possession plays (approximately 501 per match) and $ N=84432 $ direct plays (approximately 276 per match).

\begin{figure}[htbp]
  \centering
  \begin{subfigure}[t]{0.48\linewidth}
    \centering
    \includegraphics[width=\linewidth]{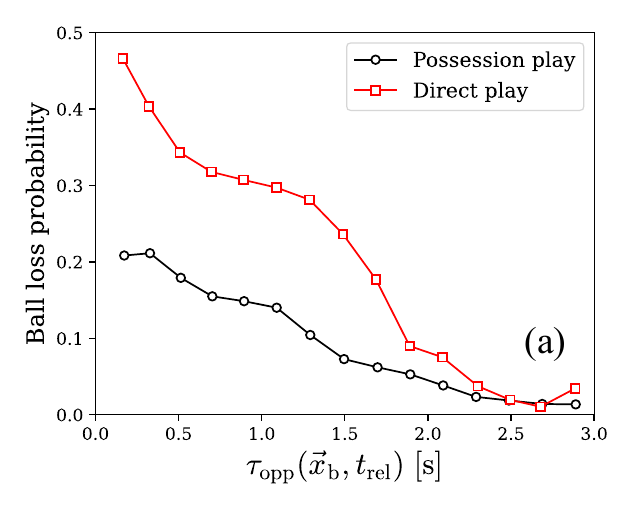}
  \end{subfigure}

  \vspace{0.7em}
  \begin{subfigure}[t]{0.48\linewidth}
    \centering
    \includegraphics[width=\linewidth]{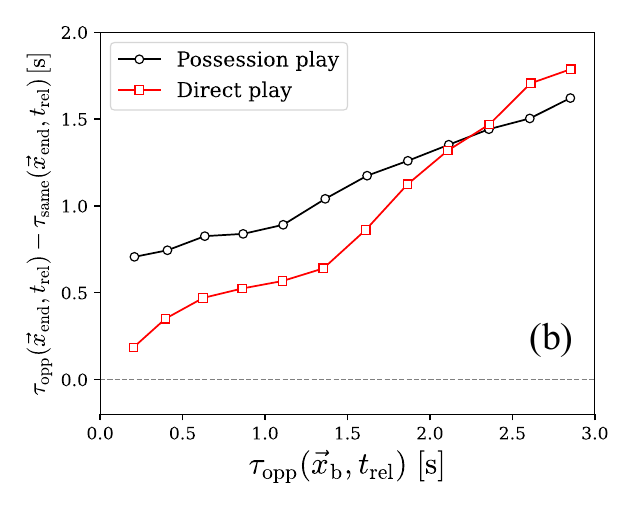}
  \end{subfigure}
  \hfill
  \begin{subfigure}[t]{0.48\linewidth}
    \centering
    \includegraphics[width=\linewidth]{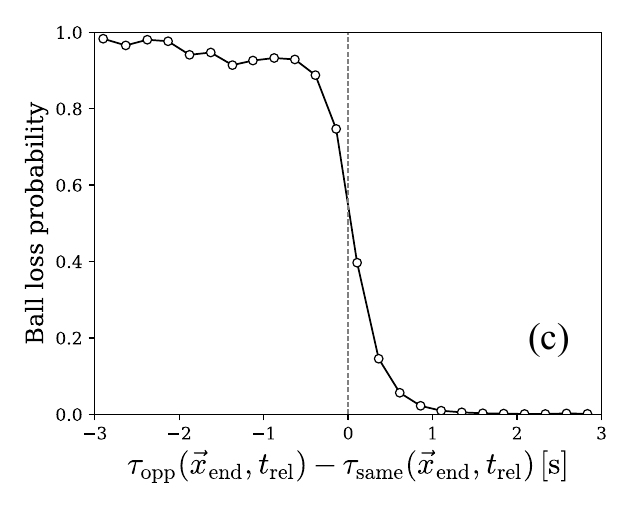}
  \end{subfigure}

  \caption{
    Association between defensive pressure at ball release, the relative safety of the realized pass-end location, and possession loss.
    (a) Ball loss probability as a function of the opponent minimum arrival time at ball release, $\tau_{\mathrm{opp}}(\vec{x}_{\mathrm{b}},t_{\mathrm{rel}})$, for possession plays (circles) and direct plays (squares).
    (b) Mean value of $\tau_{\mathrm{opp}}(\vec{x}_{\mathrm{end}}, t_{\mathrm{rel}})-\tau_{\mathrm{same}}(\vec{x}_{\mathrm{end}}, t_{\mathrm{rel}})$ as a function of $\tau_{\mathrm{opp}}(\vec{x}_{\mathrm{b}},t_{\mathrm{rel}})$, shown separately for possession plays (circles) and direct plays (squares).
    (c) Ball loss probability as a function of $\tau_{\mathrm{opp}}(\vec{x}_{\mathrm{end}}, t_{\mathrm{rel}})-\tau_{\mathrm{same}}(\vec{x}_{\mathrm{end}}, t_{\mathrm{rel}})$, pooled over all intentional open-play passes.
    Here, $\vec{x}_{\mathrm{end}}$ denotes the end location of each attempted pass.
  }
  \label{fig:loss_tau_diff}
\end{figure}

Figure~\ref{fig:loss_tau_diff}(a) shows the ball loss probability as a function of the opponent minimum arrival time at ball release, $\tau_{\mathrm{opp}}(\vec{x}_{\mathrm{b}}, t_{\mathrm{rel}})$, separately for possession plays and direct plays.
For both play types, the ball loss probability increased as $\tau_{\mathrm{opp}}(\vec{x}_{\mathrm{b}}, t_{\mathrm{rel}})$ decreased, indicating that possessions ending under stronger immediate defensive pressure were more likely to be lost.
This tendency was stronger for direct plays, whose loss probability remained higher than that of possession plays, especially when $\tau_{\mathrm{opp}}(\vec{x}_{\mathrm{b}}, t_{\mathrm{rel}})$ was small.

To interpret this tendency, we next considered the difference between the opponent and teammate minimum arrival times to the realized end location of each pass,
\[
\tau_{\mathrm{opp}}(\vec{x}_{\mathrm{end}}, t_{\mathrm{rel}})
-
\tau_{\mathrm{same}}(\vec{x}_{\mathrm{end}}, t_{\mathrm{rel}}).
\]
Here, $\vec{x}_{\mathrm{end}}$ denotes the actual end location of the attempted pass, and the arrival times are evaluated at the release time $t_{\mathrm{rel}}$.
This quantity compares how quickly opponents and teammates can reach that realized pass-end location. Smaller values indicate that teammates have less time advantage over opponents at the realized pass-end location.
Figure~\ref{fig:loss_tau_diff}(b) shows that this quantity tended to be smaller when $\tau_{\mathrm{opp}}(\vec{x}_{\mathrm{b}}, t_{\mathrm{rel}})$ was smaller, for both possession plays and direct plays.
This indicates that passes released under stronger pressure tended to end at locations where teammates had less time advantage over opponents.
For a given value of $\tau_{\mathrm{opp}}(\vec{x}_{\mathrm{b}}, t_{\mathrm{rel}})$, direct plays also tended to have smaller values than possession plays.
Thus, under comparable pressure at ball release, direct plays were associated with less time advantage for teammates over opponents at the realized pass-end location than possession plays.

We then examined whether passes ending with less time advantage for teammates over opponents were more likely to be lost. Figure~\ref{fig:loss_tau_diff}(c) shows that the ball loss probability increased monotonically as
\[
\tau_{\mathrm{opp}}(\vec{x}_{\mathrm{end}}, t_{\mathrm{rel}})
-
\tau_{\mathrm{same}}(\vec{x}_{\mathrm{end}}, t_{\mathrm{rel}})
\]
decreased.
This indicates that passes were more likely to be lost when teammates had less time advantage over opponents at the realized pass-end location.
Combined with Fig.~\ref{fig:loss_tau_diff}(b), this result supports the interpretation that passes released under stronger pressure were more likely to be lost, and that this tendency was associated with less favorable realized pass-end situations.
Since this analysis uses the realized end location $\vec{x}_{\mathrm{end}}$, it should be interpreted as a post hoc explanation of observed pass outcomes rather than as a prediction of pass risk available at the moment of ball release.
We also repeated the analysis in Fig.~\ref{fig:loss_tau_diff}(a) after dividing analyzed open-play passes into the same three field-location zones defined by the ball carrier's distance to the opponent goal at possession start.
The same qualitative relationship between \(\tau_{\mathrm{opp}}(\vec{x}_{\mathrm{b}}, t_{\mathrm{rel}})\) and ball loss probability was observed in all three zones for both possession plays and direct plays (Supplementary Fig.~S2), indicating that the association between stronger defensive pressure at ball release and higher ball loss probability is not explained solely by field location.

\section{Discussion and Conclusion}
This study examined defensive pressure on the ball carrier in soccer using the opponent minimum arrival time, $\tau_{\mathrm{opp}}(\vec{x}_{\mathrm{b}}, t)$, derived from a physics-based motion model.
Using synchronized tracking and event data from 306 J1 League matches, we analyzed both the statistical characteristics of this quantity and its temporal evolution during ball-possession intervals, and examined how it is associated with two key possession outcomes: ball progression and possession loss.
The results revealed several consistent patterns.
First, the distribution of $\tau_{\mathrm{opp}}(\vec{x}_{\mathrm{b}}, t)$ was concentrated at shorter times compared with that of $\tau_{\mathrm{same}}(\vec{x}_{\mathrm{b}}, t)$, indicating that opponents tend to reach the ball carrier more quickly than teammates.
Second, $\tau_{\mathrm{opp}}(\vec{x}_{\mathrm{b}}, t)$ typically decreased while a player was in possession and increased again at the start of the next possession after ball release.
Third, stronger defensive pressure was associated with smaller ball progression during a possession and, for intentional open-play passes, with a higher probability of losing possession at ball release.
Taken together, these results show that immediate defensive pressure on the ball carrier evolves systematically over time and is meaningfully associated with subsequent possession outcomes.
A central advantage of the present approach is its interpretability and physical grounding.
Unlike approaches that transform arrival-time information into more elaborate composite or probabilistic pressure scores, the present measure is defined directly from player kinematics through a physics-based motion model.
In particular, this quantity has a clear physical meaning as the time required for the fastest opponent to reach the ball-carrier location.
Moreover, the underlying Fujimura--Sugihara model has previously been validated under soccer players' sprint conditions using tracking data~\cite{Narizuka2023}.
These features make the opponent minimum arrival time a transparent and interpretable baseline measure of immediate defensive pressure.
Several limitations should be noted.
First, $\tau_{\mathrm{opp}}(\vec{x}_{\mathrm{b}}, t)$ is computed from a simplified motion model with fixed kinetic parameters.
Therefore, it should be interpreted as a proxy for immediate defensive pressure rather than as a complete description of future defensive behavior.
Second, the proposed metric focuses on the ball-carrier location and does not explicitly incorporate contextual factors such as body orientation, passing lanes, coordinated team movement, continuous field-position effects, or game state.
These factors may also influence ball progression and possession loss.
Although the zone-based analyses in Subsections~\ref{subsec:ball_progression} and~\ref{subsec:result_loss_analysis} suggest that the main associations are not explained solely by coarse field location, more detailed contextual modeling remains for future work.
Third, in the possession-loss analysis, the quantity
\[
\tau_{\mathrm{opp}}(\vec{x}_{\mathrm{end}}, t_{\mathrm{rel}})
-
\tau_{\mathrm{same}}(\vec{x}_{\mathrm{end}}, t_{\mathrm{rel}})
\]
was evaluated at the realized pass-end location.
Accordingly, this part of the analysis should be interpreted as a post hoc explanation of observed pass outcomes rather than as a prediction of pass risk available at the moment of ball release.
Fourth, the analysis is based on a single league and season, namely the 2023 J1 League, and the generality of the observed relationships should be examined in other competitions and tactical contexts.
One plausible interpretation of Fig.~\ref{fig:loss_tau_diff}, where increased pressure on the passer is associated with a lower time advantage at $\vec{x}_\mathrm{end}$, is that stronger immediate pressure reduces the availability of pass options or passing lanes that provide sufficient time advantage for teammates over opponents.
It is also possible that stronger pressure makes the ball carrier's pass selection or execution more difficult.
However, this post hoc analysis does not determine which explanation is more important in actual matches.
The present analysis focused on the ball-carrier location in order to quantify immediate defensive pressure in a simple and interpretable manner.
This focus also points to several natural directions for extension.
One important next step is to analyze $\tau_{\mathrm{same}}(\vec{x}_{\mathrm{b}}, t)$ as a complementary measure of local support around the ball carrier.
It will also be useful to extend the framework from the ball-carrier location to nearby space in order to evaluate passing options and local spatial structure more directly.
More broadly, incorporating contextual information such as defensive orientation, passing lanes, coordinated team movement, and game state may help connect local arrival-time-based pressure measures to broader pressing behavior at the team level.
In conclusion, this study showed that the opponent minimum arrival time provides a simple, interpretable, and physically grounded measure of immediate defensive pressure on the ball carrier.
The results indicate that this quantity captures meaningful temporal structure during possession and is associated with key possession outcomes such as ball progression and possession loss.
The proposed framework therefore provides an interpretable baseline for quantifying pressing dynamics from tracking data and for relating local defensive pressure to possession outcomes in soccer.
 \section*{Data and code availability}
The raw event and tracking data analyzed in this study are not publicly available and cannot be deposited in a public repository because they were provided by DataStadium Inc. under a data-use agreement. Anonymized and processed data, together with analysis code sufficient for verifying the analyses and reproducing the figures, are available from the corresponding author under controlled access conditions, subject to approval by DataStadium Inc. and an appropriate data-sharing agreement. Access is limited to editors, reviewers, or approved researchers for verification purposes and for a limited period; redistribution, secondary use, and transfer to third parties are not permitted.

\section*{Author contributions}
\textbf{Takuma Narizuka}: Conceptualization, Methodology, Data curation, Software, Formal analysis, Investigation, Visualization, Supervision, Funding acquisition, Writing -- original draft, Writing -- review \& editing.
\textbf{Ikuya Sakamoto}: Conceptualization, Data curation, Software, Investigation, Visualization, Writing -- original draft.
\textbf{Ken Yamamoto}: Supervision, Funding acquisition, Writing -- review \& editing.
\textbf{Yoshihiro Yamazaki}: Supervision, Writing -- review \& editing.
All authors discussed the results and approved the final manuscript.

\section*{Declaration of competing interests}
The authors declare that they have no known competing financial interests or personal relationships that could have appeared to influence the work reported in this paper.

\section*{Declaration of generative AI and AI-assisted technologies in the manuscript preparation process}
During the preparation of this work, the authors used ChatGPT to support language editing and manuscript organization.
After using this tool, the authors reviewed and edited the content as needed and take full responsibility for the content of the published article.

\section*{Acknowledgments}
The authors are grateful to DataStadium Inc., Japan, for providing the data for this study.
This work was partially supported by the Data-Centric Science Research Commons Project of the Research Organization of Information and Systems, Japan, a Grant-in-Aid for Early-Career Scientists (Grant Number JP23K16729) from the Japan Society for the Promotion of Science (JSPS), and a Grant-in-Aid for Scientific Research (C) (Grant Number JP23K03264) from JSPS.
The funders had no role in the study design, data collection and analysis, interpretation of results, preparation of the manuscript, or the decision to submit the article for publication.

\clearpage
\setcounter{figure}{0}
\setcounter{section}{0}
\renewcommand{\thefigure}{S\arabic{figure}}
\renewcommand{\thesection}{S\arabic{section}}
\begin{center}
  {\Large\bfseries Supplementary Material}
\end{center}
\vspace{1em}
\section{Zone-stratified analyses}

For the zone-stratified analyses, possessions were divided into three zones according to the ball carrier's distance to the opponent goal at possession start. The thresholds were \qty{35}{\metre} and \qty{70}{\metre}, yielding three zones: \(\mathord{<}\,\qty{35}{\metre}\), \qtyrange{35}{70}{\metre}, and \(\mathord{\ge}\,\qty{70}{\metre}\). These analyses were performed to examine whether the associations reported in the main text were explained solely by field location.

\begin{figure}[htbp]
  \centering
  \includegraphics[width=0.46\linewidth]{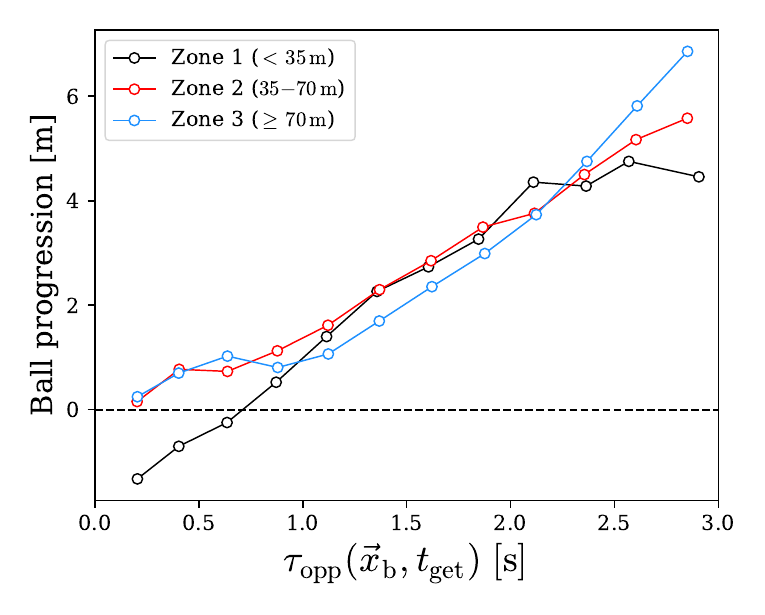}
  \caption{
    Zone-stratified association between defensive pressure at possession start and ball progression.
    Possession plays were divided into three zones according to the ball carrier's distance to the opponent goal at possession start: \(\mathord{<}\,\qty{35}{\metre}\), \qtyrange{35}{70}{\metre}, and \(\mathord{\ge}\,\qty{70}{\metre}\).
    The figure shows mean ball progression as a function of $\tau_{\mathrm{opp}}(\vec{x}_{\mathrm{b}}, t_{\mathrm{get}})$ for each zone.
  }
  \label{fig:S1_zone_ball_progression}
\end{figure}

\begin{figure}[htbp]
  \centering
  \begin{subfigure}[t]{0.46\linewidth}
    \centering
    \includegraphics[width=\linewidth]{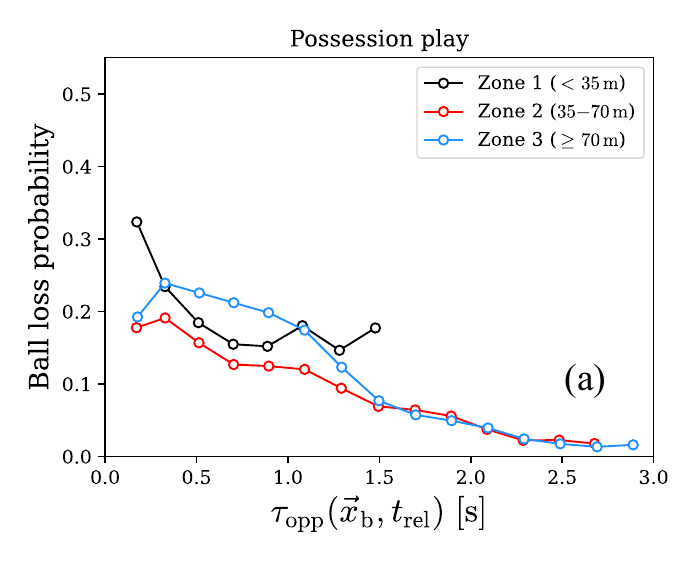}
  \end{subfigure}
  \hfill
  \begin{subfigure}[t]{0.46\linewidth}
    \centering
    \includegraphics[width=\linewidth]{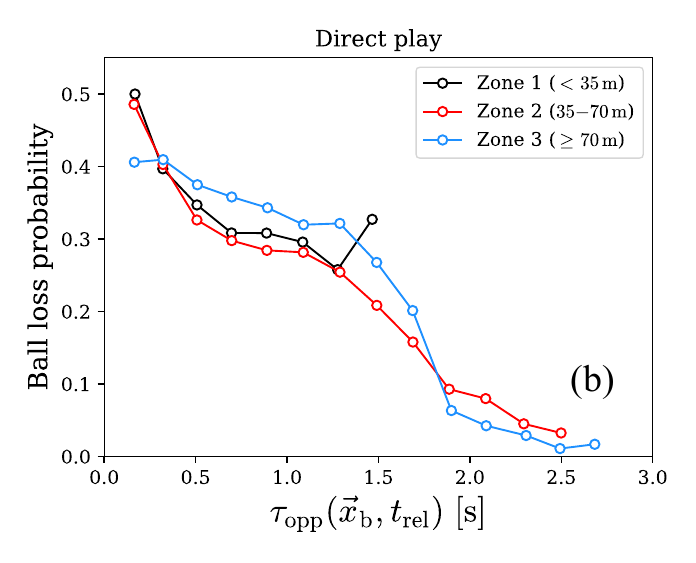}
  \end{subfigure}
  \caption{
    Zone-stratified association between defensive pressure at ball release and possession loss.
    Intentional teammate-directed open-play passes were divided into three zones according to the ball carrier's distance to the opponent goal at possession start: \(\mathord{<}\,\qty{35}{\metre}\), \qtyrange{35}{70}{\metre}, and \(\mathord{\ge}\,\qty{70}{\metre}\).
    Panels (a) and (b) show ball loss probability as a function of $\tau_{\mathrm{opp}}(\vec{x}_{\mathrm{b}}, t_{\mathrm{rel}})$ for possession plays and direct plays, respectively.
    Bins containing fewer than 50 passes were not plotted.
  }
  \label{fig:S2_zone_loss_probability}
\end{figure}

\end{document}